\documentclass{article}
\usepackage{amsmath}
\usepackage{graphicx}
\usepackage{amsfonts}
\usepackage{amssymb}
\usepackage{geometry}
\usepackage{float}
\usepackage{bm}
\usepackage{color}
\usepackage{cite}

\begin{document}

\title{Effect of Snyder-de Sitter model on the Black hole thermodynamics in
the context of rainbow gravity}
\author{B. Hamil\thanks{%
hamilbilel@gmail.com} \\
D\'{e}partement de TC de SNV, Universit\'{e} Hassiba Benbouali, Chlef,
Algeria. \and B. C. L\"{u}tf\"{u}o\u{g}lu\thanks{%
bclutfuoglu@akdeniz.edu.tr} \\
Department of Physics, Akdeniz University, Campus 07058, Antalya, Turkey, \\
Department of Physics, University of Hradec Kr\'{a}lov\'{e}, \\
Rokitansk\'{e}ho 62, 500 03 Hradec Kr\'{a}lov\'{e}, Czechia.}
\date{}
\maketitle

\begin{abstract}
One of the foremost goals of theoretical modern physics is to obtain a reliable theory of quantum gravity. In so doing, new fundamental scales are being proposed. For example,  in the Snyder - de Sitter model, two scales manifest which relates measurement limits of the position and momentum. In this work,  we study the thermodynamic functions of the Scharwzschild black hole in the Snyder - de Sitter model within the presence of  position independent and dependent gravity's rainbow. Heuristically, we prove the presence of a non-zero lower bound value of the horizon, mass, and temperature of the black hole. Then, we present the stability and remnant conditions of the black hole according to the entropy and heat capacity functions.
\end{abstract}

\section{Introduction}

\bigskip Recently, there has been growing interest in the study of classical
and quantum mechanics on a quantized spacetime \cite{1,2,3,4,5,6,7,8,9,10, c1, c2, c3, c4, c5, c6, c7}.
Snyder model \cite{11} holds an significance historical place since it is
the first model of quantum mechanics on a noncommutative spacetime which is
invariant under Lorentz transformation. In his model, based on the fundamental length assumption, Snyder redefined the
momentum operators and identified the spacetime operators as 4-`translation'
generators of de Sitter (dS) algebra within projective geometry approach. He hoped that the insertion of a fundamental length scale in quantum
field theory will overcomes the problems of UV divergences. However, the obstacle of the UV divergences in quantum field theory obtained a
solution due to the development of renormalization theory in the late forties, and then, Snyder's work was not further examined with the exception of
certain papers \cite{13,14,15,16,17,18}.

In the relativistic Snyder model the deformed Heisenberg algebra in (1+3) dimensional spacetime is defined by \cite{1,2}
\begin{equation}
\left[ X_{\mu },P_{\nu }\right] =i\hbar \left( \eta _{\mu \nu }+\beta
^{2}P_{\mu }P_{\nu }\right); \quad \quad \left[ X_{\mu },X_{\nu }\right]
=i\hbar \beta ^{2}J_{\mu \nu };  \quad \quad  \left[ P_{\mu },P_{\nu }\right] =0;
\end{equation}
where $\eta _{\mu \nu }$ is the metric tensor with the signature $\  \left[ \eta _{\mu \nu }%
\right] =\mbox{diag}\left(
\begin{array}{cccc}
-1 & 1 & 1 & 1%
\end{array}%
\right) $. $\beta $ is a coupling constant of order $1/m_{P}$ where $m_{P}$
represents the Planck mass. $J_{\mu \nu }$ are the generators of Lorentz
transformations where the spacetime indices are considered by $\mu ,\nu =0,1,2,3$. In
particular, the case $\beta ^{2}>0$ is called as the Snyder model, while $\beta
^{2}<0$ is named as the  anti-Snyder model \cite{2}.

A great deal of effort has been devoted to extending the Snyder model to spacetimes of constant curvature, by introducing a new fundamental constant proportional to the cosmological constant \cite{13}. Researches in loop quantum gravity give some justification that the cosmological constant is a necessary ingredient of the theory \cite{19}. This fact gives rise to the idea of a further generalization of the commutation relations according to this constant. From that perspective, one can introduce a third fundamental scale, $\alpha$, in addition to the speed of light $c$ and the Snyder parameter $\beta $. In this case, the model is called the Snyder-de Sitter (SdS) or triply special relativity model. Recently, several papers have been devoted to study of the effects of the
deformed canonical commutation relations on the classical and quantum systems
\cite{23,24,25,26,27,28,29,30,32,33,34,35,36,37,38}.

Rainbow gravity (RG) is one of the effective quantum gravity theories \cite{Gambini} that can be regarded as an example of doubly special relativity \cite{jMagueijo}. RG is originated back to the modification of the usual energy-momentum dispersion relation at the Planck scale \cite{Ali2015}.
\begin{eqnarray}
E^{2}f^{2}\left( \frac{E}{E_{P}%
}\right) = p^{2}c^{2}g^{2}\left( \frac{E}{E_{P}}\right) +m^{2}c^{4},
\end{eqnarray}
where $f\left( \frac{E}{E_{P}}\right) $ and $g\left( \frac{E}{E_{P}}\right) $ are two general RG functions. In gravity's rainbow, a rainbow metric is considered by the determination of the spacetime via the ratio of the energy to a test particle to $E_P$ \cite{Deng2017}.  Beside, energy can depend on position or time \cite{Momeni}. On the other hand, black holes and its thermodynamics are one of the most challenging  research topics of physics. Some of the recent studies on the field is  going on in the framework of RG \cite{BHT01, BHT02, BHT04, BHT06, BHT08, Bakke}. In this work our main motivation is to investigate the Schwarzschild black hole's thermodynamics in the SdS spacetime in the context of the gravity's rainbow to provide a contribution to the field.

We present our work as follows: In sec. \ref{sec:SdS}, we briefly introduce the SdS model and express the modification on the usual algebra and uncertainty relation. In sec. \ref{sec:BHTermho}, we choose a particular pair of RG functions after giving the most general form of the metric. Then, we obtain the mass-temperature, heat capacity and entropy functions of the black hole. In sec. \ref{sec:add}, we consider the case where the energy is considered with non homogeneous function of the coordinates. In the final section, we conclude the manuscript.

\section{Snyder-de Sitter model} \label{sec:SdS}

Let us start by considering the SdS algebra such that the commutator between the operators of the position $X_{j}$ and the momentum $%
P_{k}$ is given by \cite{Smolin,Mignemi1,Mignemi2,Mignemi3,Mignemi4}:
\begin{eqnarray}
\left[ X_{j};P_{k}\right] &=&i\hbar \Big( \eta _{jk}+\alpha
^{2}X_{j}X_{k}+\beta ^{2}P_{j}P_{k}+\alpha \beta \left(
P_{j}X_{k}+X_{k}P_{j}\right) \Big) ,  \notag \\
\left[ X_{j};X_{k}\right] &=&i\beta ^{2}\hbar \varepsilon _{jkl}L_{l};\text{
\  \  \  \  \  \  \  \  \  \  \  \ }\left[ P_{j};P_{k}\right] =i\alpha ^{2}\hbar
\varepsilon _{jkl}L_{l},  \label{2}
\end{eqnarray}
where $L_{l}$ are components of angular momentum operator. It is worth noting that in the particular values of the deformation parameter values the SdS algebra reduces to the other algebras. For example, if one considers $\alpha \rightarrow 0 $, then the SdS algebra reduces to the flat space Snyder model algebra. Alike,  in the  $\beta \rightarrow 0 $ limit, the SdS algebra yields the de Sitter algebra.  Moreover,  $\alpha, \beta \rightarrow 0$ gives the usual quantum mechanic algebra \cite{23,24}. We prefer to handle the study in the momentum representation where the position and momentum operators are given as \cite{Mignemi1} :

\begin{equation}
X_{j}=\mathcal{X}_{j}+\lambda \frac{\beta}{\alpha }\mathcal{P}_{j}=i\hbar
\sqrt{1-\beta ^{2}p^{2}}\frac{\partial }{\partial p_{j}}+\lambda \frac{%
\beta }{\alpha }\frac{p_{j}}{\sqrt{1-\beta ^{2}p^{2}}},
\end{equation}%
\begin{equation}
P_{j}=-\frac{\alpha }{\beta }\mathcal{X}_{j}+\left( 1-\lambda \right)
\mathcal{P}_{j}=-i\hbar \frac{\alpha }{\beta }\sqrt{1-\beta ^{2}p^{2}}\frac{%
\partial }{\partial p_{j}}+\left( 1-\lambda \right) \frac{p_{j}}{\sqrt{%
1-\beta ^{2}p^{2}}}.
\end{equation}
Here, $\lambda $ is an arbitrary real parameter and the range of the allowed values
of $P_{j}$ is bounded via $-\frac{1}{\beta }<P_{j}<\frac{1}{\beta }.$  This deformation in the algebra  gives rise to an uncertainty relation $%
\left( \left \langle P_{j}\right \rangle =\left \langle X_{j}\right \rangle
=0\right) $:
\begin{equation}
\left( \Delta X\right) _{j}\left( \Delta P\right) _{k}\geq \frac{\hbar }{2}%
\left( \delta _{jk}+\alpha ^{2}\left( \Delta X\right) _{j}\left( \Delta
X\right) _{k}+\beta ^{2}\left( \Delta P\right) _{j}\left( \Delta P\right)
_{k}-\alpha \beta \left( \left( \Delta P\right) _{j}\left( \Delta X\right)
_{k}+\left( \Delta X\right) _{j}\left( \Delta P\right) _{k}\right) \right) .
\label{6}
\end{equation}%
In the one-dimensional case, the uncertainty relations reduce to the following form
\begin{equation}
\left( \Delta X\right) \left( \Delta P\right) \geq \frac{\hbar }{2}\left[
1+\alpha ^{2}\left( \Delta X\right) ^{2}+\beta ^{2}\left( \Delta P\right)
^{2}-2\alpha \beta \left( \Delta X\right) \left( \Delta P\right) \right] ,
\label{SdS}
\end{equation}%
which implies non-zero minimal uncertainty values on the measurements of the position and momentum as
\begin{equation}
\left( \Delta X\right) _{\min }=\frac{\hbar \beta }{\sqrt{1+2\hbar \alpha
\beta }};\text{ \  \  \  \ }\left( \Delta P\right) _{\min }=\frac{\hbar \alpha
}{\sqrt{1+2\hbar \alpha \beta }}.
\end{equation}%
As a result of the deformation of the algebra, the definition of scalar product changes. In the SdS algebra it is defined as,
\begin{equation}
\left \langle \varphi \right. \left \vert \psi \right \rangle =\int \frac{%
d^{3}p}{\sqrt{1-\beta ^{2}p^{2}}}\varphi ^{\ast }\left( p\right) \psi
\left( p\right) .
\end{equation}
Before we proceed to the next section, we would like to note that for $\alpha ^{2},$ $\beta ^{2}<0$  minimal uncertainties do not emerge and the restriction on the $P_{i}$ changes. In this case, $P_{i}$ can take all real values.

\section{Schwarzschild black hole thermodynamics} \label{sec:BHTermho}

In this section, we explore the Schwarzschild black hole thermodynamics in the SdS model. Within the context of rainbow gravity we consider the metric of the Schwarzschild black hole in the following form \cite%
{mandal,park}
\begin{equation}
ds^{2}=-\frac{1}{f^{2}\left( \frac{E}{E_{P}}\right) }\left( 1-\frac{2MG}{%
rc^{2}}\right) c^{2}dt^{2}+\frac{1}{g^{2}\left( \frac{E}{E_{P}}\right) }%
\left( 1-\frac{2MG}{rc^{2}}\right) ^{-1}dr^{2}+\frac{r^{2}}{g^{2}\left(
\frac{E}{E_{P}}\right) }d\Omega ^{2},
\end{equation}
where $M$ represents the mass of the black hole and $G$ denotes the Newton
universal gravitational constant. It is a very well-known fact that the Schwarzschild horizon is obtained
by solving $1-\frac{2MG}{rc^{2}}=0.$ This yields to
\begin{equation}
r_{S}=\frac{2MG}{c^{2}}.  \label{9}
\end{equation}%
In this manuscript, we consider one of the most interesting forms of the rainbow functions studied in  \cite%
{Amelino1,Amelino2,Ali,Kim,Gim,Carvalho,Yadav,Tao}
\begin{equation}
f\left( \frac{E}{E_{P}}\right) =1;\quad g\left( \frac{E}{E_{P}}\right) =\sqrt{%
1-\eta \left( \frac{E}{E_{P}}\right) ^{n}}.
\end{equation}
We note that this form of RG functions is compatible with some results from the string theory \cite{Ali},  the loop quantum
gravity \cite{Amelino3} and the $\kappa $-Minkowski noncommutative spacetime
\cite{Ruegg}.  The choice of $f\left( \frac{E}{E_{P}}\right) =1$ yields a time-like Killing vector in the rainbow gravity as usual, so that, the local thermodynamic energy becomes independent from the test particle's energy.

Near-horizon geometry considerations suggests to set $\Delta X\simeq r_{S}$. Therefore,  Eq. (\ref{9}) leads to a black hole minimum value for horizon radius and mass expressions as follows:%
\begin{equation}
r_{S}\simeq \frac{\hbar \beta }{\sqrt{1+2\hbar \alpha \beta }},\quad \quad
M_{\min }=\frac{m_{P}^{2}c\beta }{2\sqrt{1+2\hbar \alpha \beta }}.\text{\ }
\end{equation}%
At the first sight, we notice that in the limit $\beta \rightarrow 0$, there is no lower bound on the black hole mass. This fact shows that a lower bound on the black hole mass owes its origin to the Snyder  model.

On the other hand,  we know that the temperature of any massless quantum particle near the Schwarzschild black hole horizon is  expressed by  \cite%
{Adler,Scardigli,mandal,park}
\begin{equation}
T=\frac{E}{K_{B}}\simeq \frac{P c }{K_{B}}\frac{g\left( \frac{E}{E_{P}}\right)
}{f\left( \frac{E}{E_{P}}\right) }. \label{12}
\end{equation}%
In the presence of minimal uncertainty in momentum, we employ the modified dispersion relation to derive a lower bound for the black hole temperature for $n=2$. We find
\begin{equation}
T_{\min }=\frac{\hbar c\alpha }{K_{B}}\frac{1}{\sqrt{1+2\hbar \alpha \beta
+\eta \left(\frac{\hbar c\alpha}{K_{B}T_{P}}\right)^2}}.
\end{equation}
In the limit of $\beta=0$, we observe a non-zero minimum temperature value
\begin{eqnarray}
T_{\min }
=\frac{T_P}{\sqrt{
\left(\frac{K_{B}T_{P}}{\hbar c \alpha }\right)^2+\eta }},
\end{eqnarray}
however,  in the limit of $\alpha = 0$, we find that there is no lower bound on the temperature. This result implies that a lower bound on the temperature is based on the de Sitter spacetime algebra. For being more concrete, we combine our conclusions and present them in Table \ref{Tab1}.
\begin{table}[htbp!]
\centering
\begin{tabular}{ | c | c | c | c | c | }
\hline
   Models               &  Parameters     & $r>r_{\min}$ & $M>M_{\min}$ & $T>T_{\min}$  \\
\hline
SdS   & $(\alpha \neq 0, \beta \neq 0)$   & \checkmark & \checkmark & \checkmark \\
\hline
Snyder &  $(\alpha = 0, \beta \neq 0)$   &\checkmark & \checkmark & $-$ \\
\hline
de Sitter  & $(\alpha \neq 0, \beta = 0)$   & $-$ & $-$ & \checkmark \\
\hline
\end{tabular} \label{Tab1}
\caption{The prediction of a minimal value for the radius, mass and temperature of the black hole.}
\end{table}

Next, we give our attention to the investigation of the effects of the SdS model on the thermodynamic properties of Schwarzschild black hole in the chosen rainbow gravity approach. Following the heuristic argument of Bekenstein \cite{Adler, Adler2001}, we write
\begin{equation}
\Delta X=\epsilon r_{S}=\frac{2\epsilon MG}{c^{2}},  \label{14}
\end{equation}
where $\epsilon $ denotes a calibration factor. By taking $\Delta P $ and $\Delta X $ into consideration from Eqs. \eqref{12} and \eqref{14},  we obtain a relation between the black hole mass and the temperature  out of  Eq. \eqref{SdS} as
\begin{equation} \label{MMM}
M=\frac{cK_{B}T}{2\epsilon G\hbar \alpha ^{2}\sqrt{1-\eta \left( \frac{T}{T_{P}} \right) ^{2}}}\left \{ 1+\hbar \alpha \beta -\sqrt{1-\left[ \left(
\frac{\hbar c\alpha }{K_{B}T}\right) ^{2}\left( 1-\eta \left( \frac{T}{T_{P}}%
\right) ^{2}\right) -2\hbar \alpha \beta \right] }\right \} .
\end{equation}%
Then, we assume $\left[ \left(
\frac{\hbar c\alpha }{K_{B}T}\right) ^{2}\left( 1-\eta \left( \frac{T}{T_{P}}%
\right) ^{2}\right) -2\hbar \alpha \beta \right]<<1$ and expand the square root term to its Taylor series up to the
second-order terms by neglecting the higher-order ones. We arrive at
\begin{equation}
M_{SdS}\simeq\frac{\left( m_{P}c\right) ^{2}}{8\pi \left( K_{B}T\right) }\sqrt{%
1-\eta \left( \frac{T}{T_{P}}\right) ^{2}}\left \{ 1-\hbar \alpha \beta +%
\frac{\left(\frac{\beta K_{B}T }{c} \right)^{2}}{\left[1-\eta \left(
\frac{T}{T_{P}}\right) ^{2}\right]}+\left( \frac{\hbar c\alpha }{2K_{B}T}\right)
^{2}\left[ 1-\eta \left( \frac{T}{T_{P}}\right) ^{2}\right] \right\}.
\label{16}
\end{equation}
Here, it is worth noting that we find a compulsory condition to get a real-valued black hole mass: $1-\eta \left( \frac{T}{T_{P}}\right) ^{2}>0$. This requirement implies that in the rainbow gravity approach the temperature is bounded not only from below, but also from above,  $T_{\min
}\leqslant T<\frac{T_{P}}{\sqrt{\eta }}$. In the absence of rainbow gravity,  for $\eta = 0$, there is no upper bound value of the temperature. This means that the presence of an upper bound temperature value owes its origin to the rainbow gravity.

\begin{figure}[htb]
\centering
\includegraphics[scale=1]{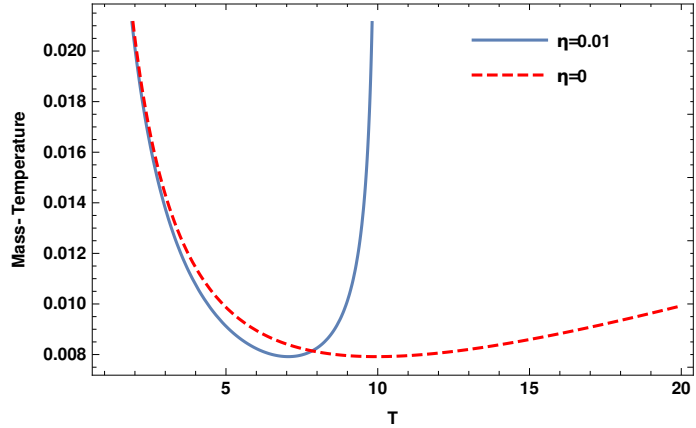}
\caption{A comparison of the mass-temperature functions in the SdS model with $\alpha=0.1$, $\beta=0.1$.}
\label{fig1}
\end{figure}
At this point, we have to mention that we are supporting our results with graphical demonstrations throughout the manuscript. Hereafter, in all figures we employ the following constants: $\hbar=c=k_B=m_P=T_P=1$. Moreover, we take the nonzero values of $\alpha$, $\beta$ and $\eta$ parameters as $0.1$, $0.1$, and $0.01$ respectively.

Initially, we plot the mass-temperature function in the SdS model versus the temperature in Fig. \ref{fig1}. We recall the condition that is required for the Taylor series expansion and use the starting point of the temperature interval much much bigger than the minimum temperature value, $0.099$, which is calculated for the chosen constants above. We observe that in the rainbow gravity, temperature's maximum value tends to $10$, which is in a complete agreement with our finding $\frac{T_{P}}{\sqrt{\eta }}$. We observe that the mass-temperature function has a non-zero minimum value. We find a minimum value for the mass-temperature function absence of the rainbow gravity. Moreover, in the  $\eta=0$ case, the temperature is limited only from below.

Within this study, we interpreter the achieved results at three critical values. At first, we consider $ \alpha=0 $  value where the discussion reduces to the flat Snyder model. Then, we take $\beta=0$ to review the results in the de Sitter space-time only. Finally, we assume $\alpha=\beta=0$ to examine the findings in flat space-time. In each case, we compare the results within the absence and presence of rainbow gravity.

\begin{figure}[htb]
\centering
\includegraphics[scale=1]{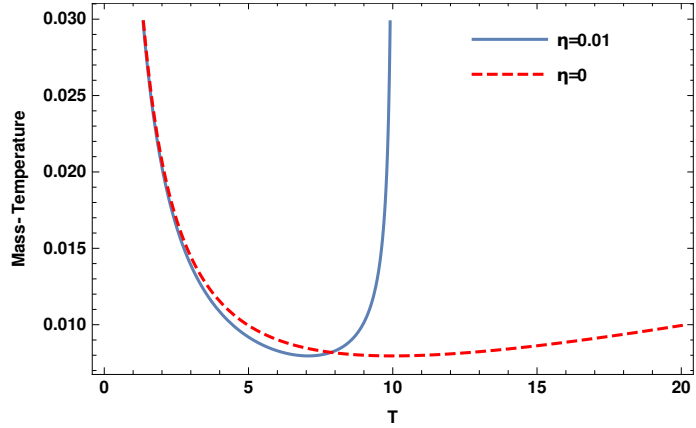}
\caption{A comparison of the mass-temperature functions in the flat Snyder model.}
\label{fig2}
\end{figure}

At first, in the framework of rainbow gravity, the mass temperature function in the flat Snyder model becomes
\begin{equation}
M_{Snyder}=\frac{\left( m_{P}c\right) ^{2}}{8\pi \left( K_{B}T\right) }\sqrt{%
1-\eta \left( \frac{T}{T_{P}}\right) ^{2}}\left( 1+\frac{\left(\frac{\beta K_{B}T }{c} \right)^{2}}{1-\eta \left(
\frac{T}{T_{P}}\right) ^{2}}\right).
  \label{19}
\end{equation}%
For $\eta =0,$ the black hole mass function reduces to
\begin{equation}
M_{Snyder}=\frac{\left( m_{P}c\right) ^{2}}{8\pi \left( K_{B}T\right) }%
\left( 1+\left(\frac{\beta K_{B}T }{c} \right)^{2}\right). \label{19b}
\end{equation}%
This result is in agreement with \cite{Hassan}. We use Eq. \eqref{19} and depict the mass-temperature function in Fig. \ref{fig2}.
We observe that the mass-temperature function in the flat Snyder model mimics the characteristic behavior of the SdS model, which confirms our prediction that a minimum value entity in the mass function is based on the Snyder model.

\begin{figure}[htb]
\centering
\includegraphics[scale=1]{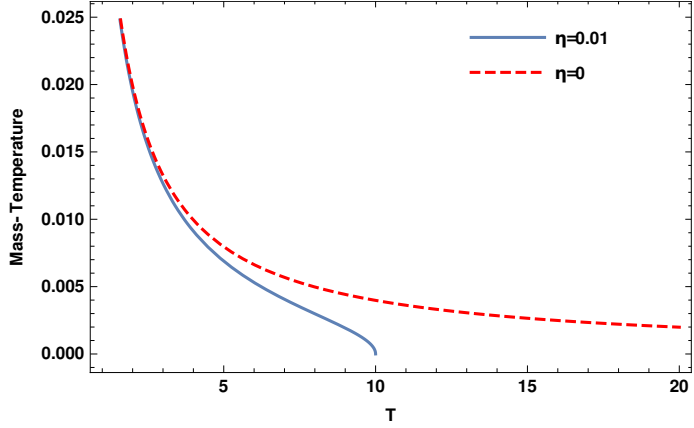}
\caption{A comparison of the mass-temperature functions in the de Sitter spacetime.}
\label{fig3}
\end{figure}

Then, in the limit $\beta =0,$ we obtain the mass-temperature function in the de Sitter
spacetime within the context of rainbow gravity,
\begin{equation} \label{MTdSR}
M_{dS}=\frac{\left( m_{P}c\right) ^{2}}{8\pi \left( K_{B}T\right) }\sqrt{%
1-\eta \left( \frac{T}{T_{P}}\right) ^{2}}\left \{ 1+\left( \frac{\hbar
c\alpha }{2K_{B}T}\right) ^{2}\left( 1-\eta \left( \frac{T}{T_{P}}\right)
^{2}\right) \right \}.
\end{equation}%
Then, we employ $\eta =0,$ to take off the rainbow gravity approach.
\begin{equation} \label{MTdS}
M_{dS}=\frac{\left( m_{P}c\right) ^{2}}{8\pi \left( K_{B}T\right) }\left(
1+\left( \frac{\hbar c\alpha }{2K_{B}T}\right) ^{2}\right) ,
\end{equation}%
When we compare our findings with the existing ones in the literature, we see that Eq. \eqref{MTdS} is in a complete agreement with the results given by Chung and Hassanabadi in \cite{Hassan}. Before we proceed to the next case, we present the plots of Eqs. \eqref{MTdSR} and \eqref{MTdS} versus temperature in Fig. \ref{fig3}.

We observe that the mass-temperature function is not minimized at a particular value other than zero. This result confirms the given prediction in Table \ref{Tab1}. Moreover, we find out that in the rainbow gravity context the mass temperature function is restricted with a maximal temperature value.

\begin{figure}[htb]
\centering
\includegraphics[scale=1]{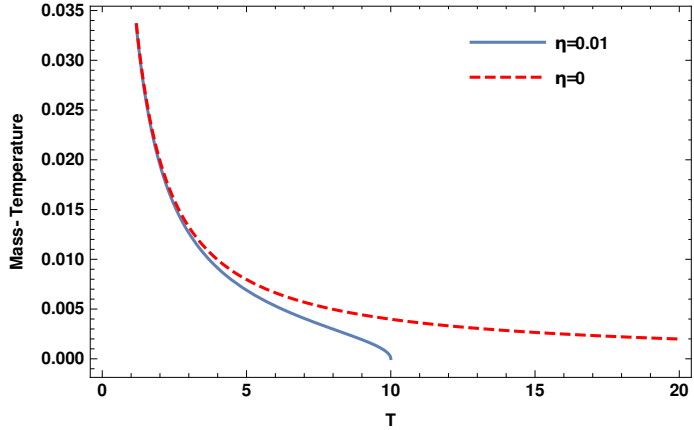}
\caption{A comparison of the mass-temperature functions in the Minkowski spacetime.}
\label{fig4}
\end{figure}

Next, in the flat space time limit, $\beta =\alpha= 0$, we find the mass-temperature function within the context of rainbow gravity in the form of%
\begin{equation}
M_{H}=\frac{\left( m_{P}c\right) ^{2}}{8\pi \left( K_{B}T\right) }%
\sqrt{1-\eta \left( \frac{T}{T_{P}}\right) ^{2}}.
\end{equation}
For $\eta =0$ rainbow gravity effect vanishes and mass-temperature function takes the form below
\begin{equation}
M_{H}=\frac{\left( m_{P}c\right) ^{2}}{8\pi \left( K_{B}T\right) },
\end{equation}
which is presented in the earlier works in \cite{Hawking,Bekenstein1,Bekenstein2}. We illustrate the behavior of the mass-temperature functions in Fig. \ref{fig4}.

We see that the mass functions display similar characteristics in the flat and in the curved spacetimes with constant curvature. In the context of rainbow gravity, it is well-defined until a particular temperature value. However, in the classical approach, such a limit does not exist.

Next, we investigate the heat capacity function of the Schwarzschild black hole in the rainbow gravity context. In the derivation we employ the following standard definition of the heat capacity
\begin{equation} \label{SH}
C=c^{2}\frac{dM}{dT}.
\end{equation}%
Then, we use Eq. \eqref{16} in Eq. \eqref{SH}. We find the specific heat in the SdS model as follows
\begin{equation}
C_{SdS}=-\frac{\left( m_{P}c^{2}\right) ^{2}}{8\pi \left( K_{B}T^{2}\right)
\sqrt{1-\eta \left( \frac{T}{T_{P}}\right) ^{2}}}\left[ 1-\hbar \alpha \beta
-\frac{\left(\frac{\beta K_{B}T }{c} \right)^{2}}{1-\eta \left(
\frac{T}{T_{P}}\right) ^{2}}+3\left( \frac{\hbar c\alpha }{2 K_{B}T}%
\right) ^{2}\left( 1-\eta \left( \frac{T}{T_{P}}\right) ^{2}\right) \right] .
\label{heat1}
\end{equation}
As in the case of the mass-temperature function, a well-defined and real-valued specific heat function is obtained under the same necessary condition.
As a natural consequence of this condition, there is an upper limit for temperature that depends only on $\eta$.

\begin{figure}[htb]
\centering
\includegraphics[scale=1]{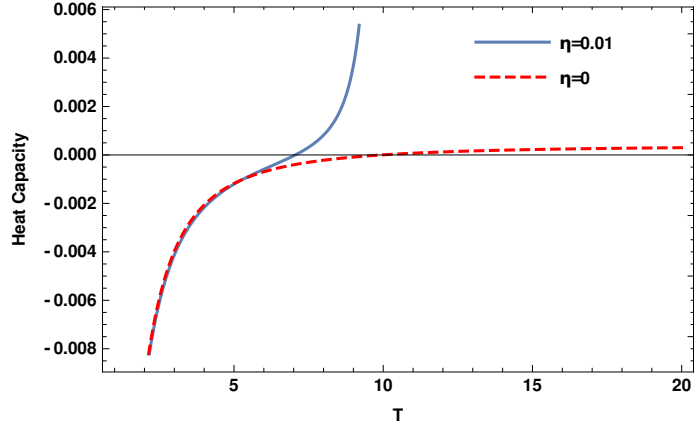}
\caption{A comparison of the mass-temperature functions in the SdS model.}
\label{fig5}
\end{figure}

We present the plot of the heat capacity function in Fig. \ref{fig5}. We observe that the heat capacity functions of the black hole become equal to zero at certain temperature values. Considering its definition, we realize that these temperature values, so called the remnant temperatures, correspond to the extreme points of the mass-temperature function. To determine these particular temperature values, we solve
\begin{eqnarray}
A+B\frac{\left( \frac{T}{T_{P}}\right) ^{2}}{1-\eta \left(
\frac{T}{T_{P}}\right) ^{2}}+D \frac{ 1-\eta \left( \frac{T}{T_{P}}\right) ^{2}}{\left( \frac{T}{T_{P}}\right) ^{2}}&=&0,
\end{eqnarray}
where
\begin{eqnarray}\label{ABC}
A=1-\hbar \alpha \beta, \quad\quad
B=-\left(\frac{\beta K_{B} T_P}{c}\right)^2, \quad \quad
D=3\left( \frac{\hbar c\alpha}{2K_{B}T_P}\right)^{2}.
\end{eqnarray}
We find the roots of the fourth order equation as follows
\begin{eqnarray}
\frac{T}{T_{P}} \label{Troots}
&=&\mp\left[\frac{\big(2 \eta D-A\big)\mp\sqrt{A^2-4BD}}{2\big(\eta^2 D-\eta A + B\big)}\right]^{1/2}.
\end{eqnarray}
We note that physically meaningful temperature root has to be real and positively valued, so that we ignore the negative root solution. Moreover, we obtain another requirement as
\begin{eqnarray}
A^2-4BD &=& 1-(2\hbar \alpha \beta) + (2\hbar \alpha \beta)^2>0.
\end{eqnarray}
It is worth noting that the square root term  does not depend on the rainbow parameter. For $\eta=0$, the solution reduces to
\begin{eqnarray}
\frac{T}{T_{P}} &=& \left[-\frac{A }{2 B}\left(1\mp\sqrt{1-\frac{4BD}{A^2}}\right)\right]^{1/2}.
\end{eqnarray}
while
\begin{eqnarray} \label{a2b}
 -\frac{A}{2B}&=& \frac{1-\hbar \alpha \beta}{\left(\frac{\beta K_{B} T_P}{c}\right)^2}.
\end{eqnarray}
For a nonzero  value of $ \beta $ parameter, Eq. \eqref{a2b} is always going to yield a non-zero value, except for $\alpha =1/(\hbar \beta) $.
Therefore, we conclude that a  remnant temperature value owes its back only to the Snyder model.

In the SdS model, for the chosen parameters above, we calculate the remnant temperature values. In the presence of the rainbow gravity, we find it as $7.05$, while in the absence it becomes to $9.95$. Then,  with these remnant temperatures we obtain the minimum mass-temperature values as $7.92 \times 10^{-3}$ from Eqs. \eqref{19} and \eqref{19b}.
It is worth noting that the plots in Fig. \ref{fig1} and Fig. \ref{fig5} confirm these results.

\begin{figure}[htb]
\centering
\includegraphics[scale=1]{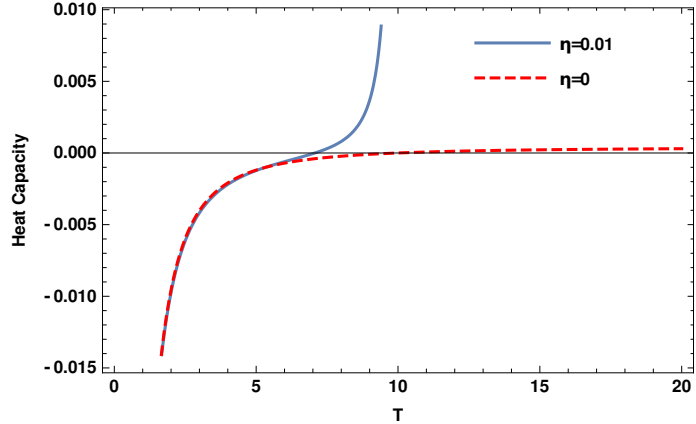}
\caption{A comparison of the mass-temperature functions in the flat Snyder model.}
\label{fig6}
\end{figure}

In the flat Snyder model,  of the terms expressed in the Eq. \eqref{ABC},  $B$ remains the same, however $A$ and $D$ turn to $1$ and $0$, respectively. Therefore, we get a non zero and positive remnant temperature expression in the form of
\begin{eqnarray}
T^{Rem}_{Sny}&=&\frac{T_P}{\sqrt{\eta+\left(\frac{\beta K_{B} T_P}{c}\right)^2}}.\label{25}
\end{eqnarray}
According to the chosen parameters above, we evaluate the remnant temperature value as $7.07$ and $10.0$ in the presence and absence of rainbow gravity. At these temperatures, the mass-temperature function reaches its minimum value that is calculated to be equal to $7.96 \times 10^{-3}$. Alike in the SdS model, the minimum mass of the black hole value in the flat Snyder model is same in the presence and absence of rainbow gravity. To obtain the same minimum mass value at different approaches and different temperatures is an extremely interesting result. In order not to make a misleading judgment, we decide to derive an algebraic expression to the minimum mass of the black hole at the remnant temperature. In so doing, we substitute Eq. \eqref{25} in Eq. \eqref{19}. We find
\begin{equation}
M_{Sny}^{\min}=\beta \frac{\left( m_{P}c\right) ^{2}}{4\pi c},
\end{equation}
We find that the expression of the remnant mass does not depend on the rainbow gravity parameter $\eta $. Therefore, we conclude that our numerical findings are not spurious. Then, we derive the heat capacity of the black hole in the flat Snyder model.
\begin{equation}
C_{Sny}=-\frac{\left( m_{P}c^{2}\right) ^{2}}{8\pi \left( K_{B}T^{2}\right)
\sqrt{1-\eta \left( \frac{T}{T_{P}}\right) ^{2}}}\left[ 1-\frac{\left(\frac{\beta K_{B}T }{c} \right)^{2}}{1-\eta \left(
\frac{T}{T_{P}}\right) ^{2}} \right] \label{24}
\end{equation}%
In Fig. \ref{fig6}, we plot the heat capacity function versus the temperature.

The plots approve the same characteristic behaviors of the heat capacity function in the SdS and Snyder models with and without rainbow gravity. Moreover, we observe that heat capacity functions vanish at the remnant temperature values which are calculated above.

Next, we analyze the heat capacity function in the de Sitter spacetime solely.  In so doing, we consider the following parameters $\alpha\neq 0$ and $\beta=0$.  In this case,  we find that $A$ and $B$ become $1$ and $0$, respectively, while $D$ remains the same.  Under the condition of $\eta D> 1$, Eq. \eqref{Troots} produces two positive and non trivial roots, namely $\frac{T_P}{\sqrt{\eta}}$ and $\frac{ T_P}{\sqrt{\eta-\frac{1}{D}}}$, where $\frac{T_P}{\sqrt{\eta}}<\frac{ T_P}{\sqrt{\eta-\frac{1}{D}}}$.  For $\eta D< 1$,  the second root turns to a complex-valued number, thus, becomes unphysical. Since in the rainbow gravity context, the temperature is defined always less than $\frac{T_P}{\sqrt{\eta}}$, a remnant temperature value cannot be seen in both cases. This is the mathematical fact that underlies the interpretation that the black hole does not have a minimum mass value in the de-Sitter spacetime. It is worth noting that, in the absence of the rainbow gravity,  one of the roots goes to infinity while the other becomes unphysical.

\begin{figure}[htb]
\centering
\includegraphics[scale=1]{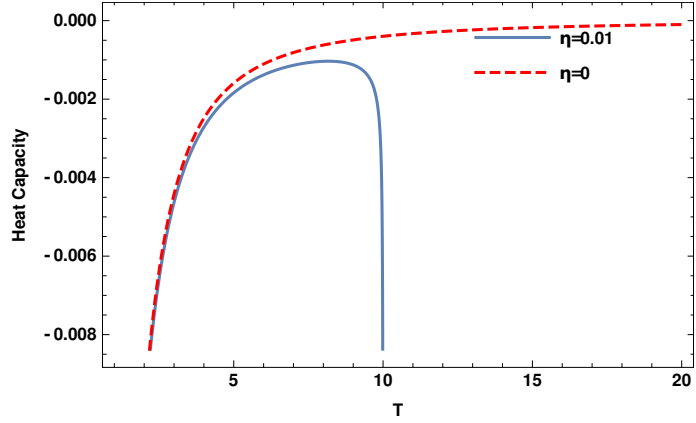}
\caption{A comparison of the  heat capacity functions in the de Sitter spacetime.}
\label{fig7}
\end{figure}

By using $\beta=0$ in Eq. \eqref{heat1}, we acquire the heat capacity function in the de-Sitter spacetime in following form
\begin{equation}
C_{dS}=-\frac{\left( m_{P}c^{2}\right) ^{2}}{8\pi \left( K_{B}T^{2}\right)}\left[\frac{1}{\left( 1-\eta \left( \frac{T}{T_{P}}\right) ^{2}\right)} +3\left( \frac{\hbar c\alpha }{2 K_{B}T}%
\right)^{2} \right] \sqrt{1-\eta \left( \frac{T}{T_{P}}\right) ^{2}} .
\end{equation}%
In Fig. \ref{fig7}, we plot the heat capacity function versus the temperature.

\begin{figure}[htb]
\centering
\includegraphics[scale=1]{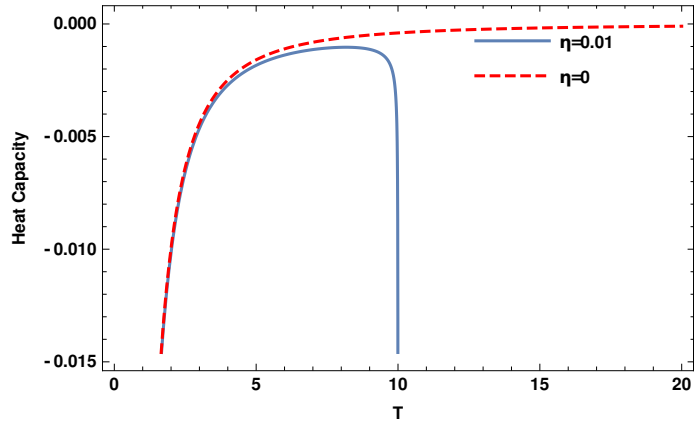}
\caption{A comparison of the heat capacity functions in the Minkowski spacetime.}
\label{fig8}
\end{figure}

Next, we consider a flat Minkowski spacetime  by selecting $ \alpha = \beta = 0 $ in the context of rainbow gravity. In this case,  $B$ and $D$ vanish while $A$ reduces to $1$. Therefore, Eq. \eqref{Troots}  yields only one real and nonzero root in the form of
\begin{eqnarray}
T^{Rem}=T_P/\sqrt{\eta}.
\end{eqnarray}
This result is the same with the de-Sitter case. This is the mathematical proof of the fact which commits  that a remnant temperature and a minimum mass does not exist in the flat or curved-spacetime with a constant curvature. We turn off the deformation parameters in Eq. \eqref{heat1}, and express the heat capacity function as

\begin{equation}
C_{H}=-\frac{\left( m_{P}c^{2}\right) ^{2}}{8\pi \left(
K_{B}T^{2}\right) \sqrt{1-\eta \left( \frac{T}{T_{P}}\right) ^{2}}}.
\end{equation}
In the limit of $\eta \rightarrow 0$, the standard heat capacity function is obtained. Before we proceed to the next thermodynamic function, we
present the plot of the heat capacity function in Fig. \ref{fig8}.

\begin{figure}[htb]
\centering
\includegraphics[scale=1]{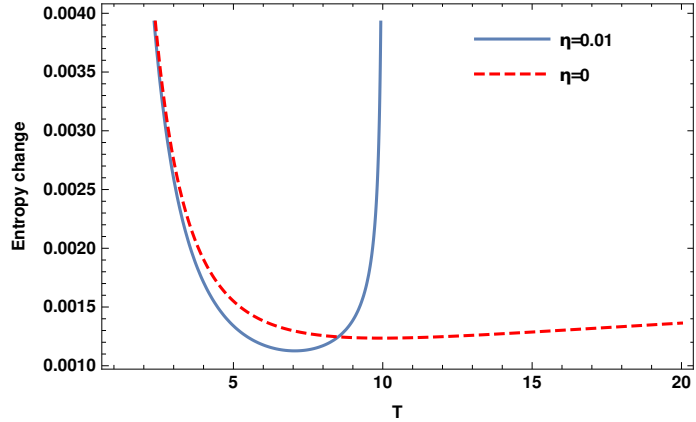}
\caption{A comparison of the entropy functions in the SdS model.}
\label{fig9}
\end{figure}

Finally, we examine another important quantity of the black hole, its entropy. In so doing we employ the well-known formula
\begin{equation}
S=\int \frac{dM}{T}=\int C\left( T\right) \frac{dT}{T}. \label{Entr}
\end{equation}%
Regarding the SdS model we substitute Eq.(\ref{heat1}) in Eq. \eqref{Entr}. We find

\begin{eqnarray}
S_{SdS}&=&S_0+\frac{\left( m_{P}c^{2}\right)^{2}}{16\pi K_{B}T_P^2}\Bigg[\Bigg(\frac{3\eta^2 }{4}\left( \frac{\hbar c \alpha }{2 K_{B}T_P}\right) ^{2}-\eta(1-\hbar\alpha\beta)+2\left(\frac{\beta K_{B}T_P }{c} \right)^{2}\Bigg)\log \frac{\left( \frac{T}{T_P}\right)}{1+\sqrt{1-\eta \left( \frac{T}{T_P}\right)^2}} \\
&+&\left(\frac{\sqrt{1-\eta \left( \frac{T}{T_P}\right)^{2}}}{\left( \frac{T}{T_P}\right)^2} \right)\Bigg(1-\hbar \alpha \beta + 2\left(\frac{\beta K_{B}T_P }{c} \right)^{2}\frac{\left( \frac{T}{T_P}\right)^2}{1-\eta \left( \frac{T}{T_P}\right)^{2}}+\frac{3}{4}\left( \frac{\hbar c \alpha }{2 K_{B}T_P}\right) ^{2} \frac{\Big(2-\eta \left( \frac{T}{T_P}\right)^2\Big)}{\left( \frac{T}{T_P}\right)^2}\Bigg)\Bigg], \nonumber  \label{ENTSdS}
\end{eqnarray}
where $S_0$ is an integration constant. We depict the entropy change in the presence and absence of the rainbow gravity in Fig. \ref{fig9}.

In the presence and in the absence of the rainbow gravity,  we observe the same characteristic behavior in the black hole's entropy. Such that, entropy increases with a decreasing increment until a certain temperature value, after that, with an increasing increment. The main difference between the two approaches is that these changes are experienced more sharply due to a certain upper temperature value in the rainbow gravity.

In the flat Snyder model, the entropy function reduces to the following form
\begin{eqnarray}
S_{Sny}&=& S_0+\frac{\left( m_{P}c^{2}\right)^{2}}{16\pi K_{B}T_P^2}\Bigg[\Bigg(2\left(\frac{\beta K_{B}T_P }{c} \right)^{2}-\eta \Bigg)\log \frac{\left( \frac{T}{T_P}\right)}{1+\sqrt{1-\eta \left( \frac{T}{T_P}\right)^2}} \nonumber \\
&+&\left(\frac{\sqrt{1-\eta \left( \frac{T}{T_P}\right)^{2}}}{\left( \frac{T}{T_P}\right)^2} \right)\Bigg(1 + 2\left(\frac{\beta K_{B}T_P }{c} \right)^{2}\frac{\left( \frac{T}{T_P}\right)^2}{1-\eta \left( \frac{T}{T_P}\right)^{2}}\Bigg)\Bigg].
\end{eqnarray}

\begin{figure}[htb]
\centering
\includegraphics[scale=1]{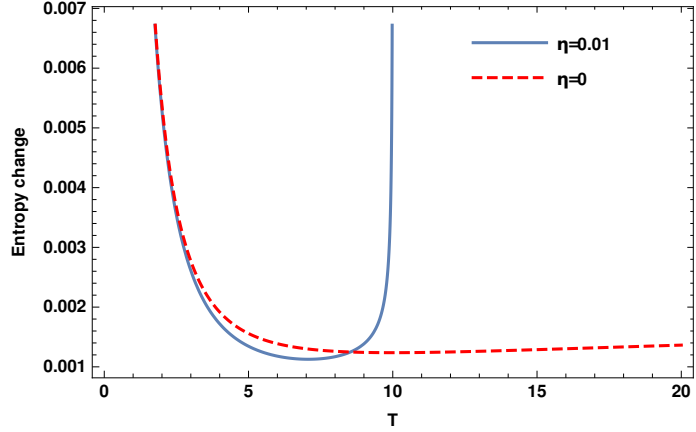}
\caption{A comparison of the entropy functions in the flat Snyder model.}
\label{fig10}
\end{figure}

\begin{figure}[htb]
\centering
\includegraphics[scale=1]{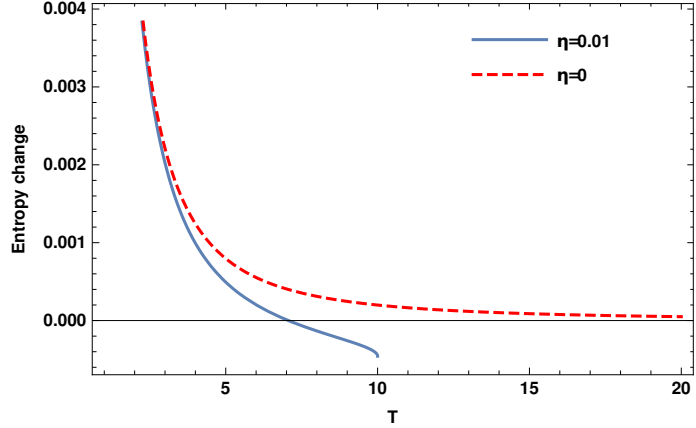}
\caption{A comparison of the entropy functions in the de Sitter spacetime.}
\label{fig11}
\end{figure}

We plot the black hole's entropy function in Fig. \ref{fig10}.
We observe that the same characteristic behavior of the SdS model. Then, we explore the entropy in the de-Sitter spacetime in the absence of the Snyder model. In this case, Eq. \eqref{ENTSdS} yields to
\begin{eqnarray}
S_{dS}&=&S_0+ \frac{\left( m_{P}c^{2}\right)^{2}}{16\pi K_{B}T_P^2}\Bigg[\Bigg(\frac{3\eta^2 }{4}\left( \frac{\hbar c \alpha }{2 K_{B}T_P}\right) ^{2}-\eta\Bigg)\log \frac{\left( \frac{T}{T_P}\right)}{1+\sqrt{1-\eta \left( \frac{T}{T_P}\right)^2}} \nonumber\\
&+&\left(\frac{\sqrt{1-\eta \left( \frac{T}{T_P}\right)^{2}}}{\left( \frac{T}{T_P}\right)^2} \right)\Bigg(1+\frac{3}{4}\left( \frac{\hbar c \alpha }{2 K_{B}T_P}\right) ^{2} \frac{\Big(2-\eta \left( \frac{T}{T_P}\right)^2\Big)}{\left( \frac{T}{T_P}\right)^2}\Bigg)\Bigg].
\end{eqnarray}
We demonstrate the entropy change in Fig. \ref{fig11}.

In the rainbow gravity, the black hole's entropy increases with a decreasing increment up to a certain temperature, however unlike the SdS and  flat Snyder models, after the critical temperature the entropy decreases until the upper temperature limit.  In the absence of the rain gravity, the entropy increase diminish at very high temperatures and the entropy function saturates at a certain entropy value.

\begin{figure}[htb]
\centering
\includegraphics[scale=1]{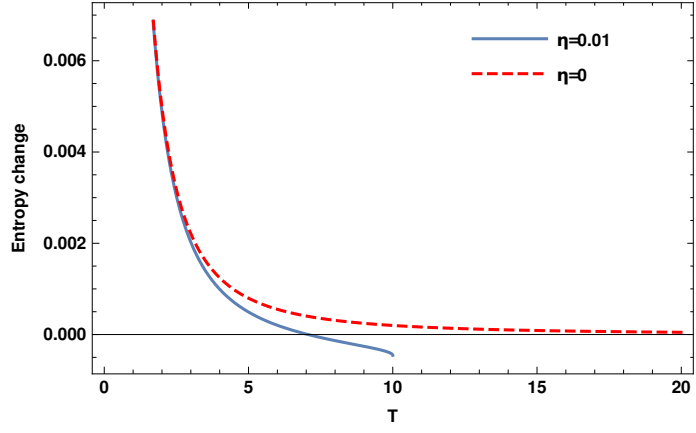}
\caption{A comparison of the entropy functions in the Minkowski spacetime.}
\label{fig12}
\end{figure}

At last, we shrink the entropy function to the context of the flat Minkowski space. When $\alpha$ and $\beta$ parameters vanish, we obtain
\begin{eqnarray}
S_{H}&=& S_0+\frac{\left( m_{P}c^{2}\right)^{2}}{16\pi K_{B}T_P^2}\Bigg[-\eta\log \frac{\left( \frac{T}{T_P}\right)}{1+\sqrt{1-\eta \left( \frac{T}{T_P}\right)^2}} +\frac{\sqrt{1-\eta \left( \frac{T}{T_P}\right)^{2}}}{\left( \frac{T}{T_P}\right)^2} \Bigg].  \label{entMin}
\end{eqnarray}
It is worth noting that in the limit $\eta \rightarrow 0,$ we recover the usual result
\begin{eqnarray}
S_{H}&=&\frac{\left( m_{P}c^{2}\right)^{2}}{16\pi K_{B}T^2}.
\end{eqnarray}
We use Eq. \eqref{entMin} to display the entropy function change versus temperature in Fig. \ref{fig12}.

When we compare Fig. \ref{fig11} and Fig. \ref{fig12}, we observe a similarity. We conclude that entropy functions present the same behavior in the flat and the curved spacetime with constant curvature.

{
\section{Coordinate dependent rainbow functions} \label{sec:add}
In this section, we are inspired from \cite{Momeni} and  investigate the special case where the energy  depends on the spatial coordinates, $E=E(r)$. To this aim, we consider the line in the form of
\begin{eqnarray}
ds^{2}&=&-\mathcal{A}\left( r\right) dt^{2}+\frac{1}{\mathcal{B}\left(
r\right) }dr^{2}+\frac{r^{2}}{g^{2}\left( \frac{E\left( r\right) }{E_{P}}%
\right) }d\Omega ^{2},
\end{eqnarray}
where%
\begin{eqnarray}
\mathcal{A}\left( r\right) &=&\frac{1}{f^{2}\left( \frac{E\left( r\right) }{%
E_{P}}\right) }\left( 1-\frac{2MG}{rc^{2}}\right), \\
\mathcal{B}%
\left( r\right) &=&g^{2}\left( \frac{E\left( r\right) }{E_{P}}\right) \left( 1-%
\frac{2MG}{rc^{2}}\right).
\end{eqnarray}
Then, we express the modified temperature %
\begin{eqnarray}
T=\frac{\kappa }{8\pi }\left( \frac{\Delta A}{\Delta S}\right) _{\min }
\end{eqnarray}
where the surface gravity can be derived from \cite{Ali2014}
\begin{eqnarray}
\kappa =\left. \sqrt{\frac{\partial \mathcal{A}\left( r\right) }{\partial r}%
\frac{\partial \mathcal{B}\left( r\right) }{\partial r}}\right \vert
_{r=r_{S}}=\frac{1}{r_{S}}\frac{g\left( \frac{E\left( r_{S}\right) }{E_{P}}%
\right) }{f\left( \frac{E\left( r_{S}\right) }{E_{P}}\right) },
\end{eqnarray}
and
\begin{eqnarray}
\left( \frac{\Delta A}{\Delta S}\right) _{\min }=\frac{\epsilon \left(
\Delta X\right) \left( \Delta P\right) }{\ln 2}.
\end{eqnarray}
Accordingly we write
\begin{eqnarray}
\left( \Delta A\right) _{\min }=\epsilon \Delta X\Delta P=\frac{\epsilon
\left( \Delta X\right) ^{2}}{\beta ^{2}\hbar }\left( 1+\hbar \alpha \beta -%
\sqrt{1+\hbar \alpha \beta -\frac{\beta ^{2}\hbar ^{2}}{\left( \Delta
X\right) ^{2}}}\right),
\end{eqnarray}
which reduces to the form of
\begin{eqnarray}
\left( \Delta A\right) _{\min }\simeq \frac{\epsilon \hbar }{2}\left[
1-\hbar \alpha \beta +\alpha ^{2}\left( \Delta X\right) ^{2}+\allowbreak
\frac{\hbar ^{2}\beta ^{2}}{4\left( \Delta X\right) ^{2}}+..\right]
\end{eqnarray}
for $\alpha <<1$ and $\beta <<1$. According to the near-horizon geometry consideration, as mentioned in the third section, we take $\left( \Delta X\right) =r_{S}$, and obtain the modified temperature as
\begin{eqnarray}
T=\frac{\epsilon \hbar }{16\pi \ln 2}\frac{1}{r_{S}}\left[ 1-\hbar \alpha
\beta +\alpha ^{2}r_{S}^{2}+\allowbreak \frac{\hbar ^{2}\beta ^{2}}{%
4r_{S}^{2}}+..\right] \frac{g\left( \frac{E\left( r_{S}\right) }{E_{P}}%
\right) }{f\left( \frac{E\left( r_{S}\right) }{E_{P}}\right) }.
\end{eqnarray}
To determine $\epsilon$, we consider $\alpha =\beta =0$ and $E_{P}\rightarrow \infty $. Therefore, we arrive at
\begin{eqnarray}
T=\frac{\hbar }{4\pi r_{S}}=\frac{\epsilon \hbar }{16\pi \ln 2}\frac{1}{r_{S}%
},
\end{eqnarray}
which leads to
\begin{eqnarray}
\epsilon =4\ln 2.
\end{eqnarray}
We conclude that for position dependent energy case, the final form of the modified temperature becomes
\begin{eqnarray}
T=\frac{\hbar }{4\pi r_{S}}\left[ 1-\hbar \alpha \beta +\alpha
^{2}r_{S}^{2}+\allowbreak \frac{\hbar ^{2}\beta ^{2}}{4r_{S}^{2}}+..\right]
\frac{g\left( \frac{E\left( r_{S}\right) }{E_{P}}\right) }{f\left( \frac{%
E\left( r_{S}\right) }{E_{P}}\right) }. \label{edept}
\end{eqnarray}
For the following particular choice of the RG functions
\begin{eqnarray}
f\left( \frac{E\left( r_{S}\right)
}{E_{P}}\right) =1 ;\quad
g\left( \frac{E\left( r_{S}\right) }{E_{P}}\right) =\sqrt{1-\eta \frac{%
E\left( r_{S}\right) }{E_{P}}}, \label{AB1}
\end{eqnarray}
Eq. \eqref{edept} reads
\begin{eqnarray}
T=\frac{\hbar }{4\pi r_{S}}\left[ 1-\hbar \alpha \beta +\alpha
^{2}r_{S}^{2}+\allowbreak \frac{\hbar ^{2}\beta ^{2}}{4r_{S}^{2}}+..\right]
\sqrt{1-\eta \frac{E\left( r_{S}\right) }{E_{P}}}.
\end{eqnarray}
Next, we derive the heat capacity function in the latter assumption. To this end we employ Eq. \eqref{SH} and after straightforward algebra we find
\begin{eqnarray}
C&=&\frac{-\frac{2\pi c^{2}}{\hbar G}f^{2}r_{S}^{2}}{\left[
1-\hbar \alpha \beta -\alpha ^{2}r_{S}^{2}+\allowbreak \frac{3\hbar
^{2}\beta ^{2}}{4r_{S}^{2}}\right] gf-r_{S}\left[ 1-\hbar \alpha \beta
+\alpha ^{2}r_{S}^{2}+\allowbreak \frac{\hbar ^{2}\beta ^{2}}{4r_{S}^{2}}%
\right] \frac{E^{\prime }\left( r_{S}\right) }{E_{P}}\left( g^{\prime }f-f^{\prime }%
g\right) }.
\end{eqnarray}
With the particular choice that is given in Eq. \eqref{AB1}, heat capacity reduces to
\begin{eqnarray}
C &=&\frac{-\frac{2\pi c^{2}}{\hbar G}r_{S}^{2}\sqrt{1-\eta \frac{E\left(
r_{S}\right) }{E_{P}}}}{\left[ 1-\hbar \alpha \beta -\alpha
^{2}r_{S}^{2}+\allowbreak \frac{3\hbar ^{2}\beta ^{2}}{4r_{S}^{2}}\right]
\left( 1-\eta \frac{E\left( r_{S}\right) }{E_{P}}\right) +\frac{\eta }{2}%
r_{S}\left[ 1-\hbar \alpha \beta +\alpha ^{2}r_{S}^{2}+\allowbreak \frac{%
\hbar ^{2}\beta ^{2}}{4r_{S}^{2}}\right] \frac{E^{\prime }\left(
r_{S}\right) }{E_{P}}}.
\end{eqnarray}
We observe that a black hole remnant occurs if and only if the following condition is satisfied: $1-\eta \frac{E\left( r_{S}\right) }{E_{P}}=0$.
Finally, we derive the entropy function via the definition given in Eq. \eqref{Entr}. We find
\begin{eqnarray}
S&=&\frac{2\pi c^{2}}{%
\hbar G}\int { \frac{r_{S}dr_{S}}{\left[ 1-\hbar \alpha \beta +\alpha
^{2}r_{S}^{2}+\allowbreak \frac{\hbar ^{2}\beta ^{2}}{4r_{S}^{2}}+..\right]
\frac{g\left( \frac{E\left( r_{S}\right) }{E_{P}}\right) }{f\left( \frac{%
E\left( r_{S}\right) }{E_{P}}\right) }}}.
\end{eqnarray}
}

\section{Conclusion}
In this manuscript, according to the SdS model we study the thermodynamic functions of the Schwarzschild black hole in the rainbow gravity. At first, we show the presence of a non-zero lower bound value of the horizon, mass, and temperature of the black hole. Then, we explore the mathematical background of these values.  We find that a non-zero minimum value of horizon radius and mass rely on the Snyder model, while the non-zero temperature depends on the de-Sitter spacetime. We obtain the mass-temperature function of the black hole and investigate it in flat Snyder, de Sitter and Minkowski space-time in the the RG formalism. Then, we derive the heat capacity function. We find that {the} black hole cannot exchange heat with the surrounding space at some critical temperatures. We execute a detailed analyze on the determination of these temperatures and black hole remnants in SdS, flat Snyder, de Sitter, and Minkowski spacetime limits. { Next,} we observe that in the SdS and Snyder models, { as well as in the} de-Sitter and Minkowski space times the thermal quantities have the same characteristic behaviors. { Finally, we consider the position dependent energy and revise the thermal quantities.} During the manuscript, we support our findings and discussion with various plots.



{
\section*{Acknowledgements}
The authors thank the anonymous reviewer for his/her helpful and constructive comments.}

\end{document}